\documentclass[12pt,english]{article}
\usepackage{babel}
\usepackage[cp1250]{inputenc}
\usepackage[T1]{fontenc}
\usepackage{amsfonts}
\usepackage{amsmath}
\usepackage{amsthm}
\usepackage{hyperref}
\usepackage{enumerate}
\usepackage{graphicx}
\usepackage{pictex}
\usepackage{color}

\newcommand{\eqq}[2]{\begin{equation}  #1  \label{#2} \end{equation}    }

\newcommand{\hd}{\hspace{0.2cm}}
\newcommand{\no}{\noindent}
\newcommand{\m}[1]{\mbox{ #1}}
\newcommand{\ww}{\subseteq }
\newcommand{\Om}{\Omega}
\newcommand{\ddt}{\frac{d}{dt}}
\newcommand{\ddta}{\frac{d}{d \tau}}

\newcommand{\rlz}{{}^{\m{RL}}{}_{0}}
\newcommand{\rls}{{}^{\m{RL} }_{s(t)}}
\newcommand{\cs}{{}^{\m{C} }_{s(t)}}

\newcommand{\dt}[1]{D_{t}^{#1}}
\newcommand{\dtb}{\dt{\beta}}
\newcommand{\dtjmb}{\dt{1- \beta}}
\newcommand{\Oml}{\Om_{l}}
\newcommand{\Omlt}{\Oml(t)}
\newcommand{\Oms}{\Om_{s}}
\newcommand{\Omst}{\Oms(t)}
\newcommand{\vlt}{V_{l}(t)}
\newcommand{\vst}{V_{s}(t)}

\newcommand{\nk}[1]{ \left[ #1 \right]}

\newcommand{\intfjmb}{{}_{0}I^{1-\beta}_{t}}
\newcommand{\cabt}{{}^{C} {}_{0}\dtb}
\newcommand{\ep}{\varepsilon}
\newcommand{\ti}[1]{\tilde{#1}}
\newcommand{\sma}{s^{-1}(a)}

\newtheorem{remark}{Remark}

\begin{document}

\title{A note about fractional Stefan problem}

\author{A. Kubica,  K. Ryszewska\\
\medskip\\
Department of Mathematics and Information Sciences\\
Warsaw University of Technology\\
pl. Politechniki 1, 00-661 Warsaw, Poland\\e-mail:
{\tt A.Kubica@mini.pw.edu.pl}\\
{\tt K.Ryszewska@mini.pw.edu.pl}
}

\maketitle
\begin{quote}
 \footnotesize
{\bf Abstract} We derive the fractional version of one-phase one-dimensional Stefan model. We assume that the diffusive flux is given by the time-fractional Riemann-Liouville derivative, i.e. we impose the memory effect in the examined model.
\end{quote}

\bigskip\noindent
{\bf Key words:} fractional derivatives, Stefan problem.

\bigskip\noindent
{\bf 2010 Mathematics Subject Classification.} Primary: 35R11 Secondary: 35R37

\section{Introduction}
The purpose of this paper is to derive the
one-dimensional one-phase Stefan model. We are motivated by the paper \cite{FV}, where the authors represent the nonlocal in time effects by assuming that the diffusive flux is given in the form of time-fractional Riemann-Liouville derivative of temperature gradient. The similar result has been already obtained in \cite{Ros}.

\section{Formulation of the problem}
We assume that $\Om=(0,L)$ for a positive $L$. Further, we assume that at the initial time $t=0$ the domain $\Om$ is divided onto two parts: $(0,x_{0})$ - ``liquid'' and $(x_{0},L)$ - ``solid''. In particular, we admit the case where $x_{0}=0$. Following \cite{FV} we define the enthalpy function by $E= T+\phi$, where $T(x,t)$ is the temperature at point $x\in \Om$ at time $t$ and $\phi$ represents the latent heat.
We consider the sharp-interface model, hence we assume that $\phi$ is given in the following form
\eqq{\phi = \left\{ \begin{array}{ll}  1 & \m{ in liquid }  \\
0  & \m{ in solid. }  \\ \end{array}  \right.}{a1}
We shall consider the one-phase model, i.e. we assume that $T\equiv 0$ in ``solid'' part. We denote by $q^{*}(x,t)$ the flux at $x\in \Om$ at time $t$. The main principle, which we assume is the conservation law which here has the following form: for each $V=(a,b)\ww \Om$
\eqq{\ddt \int_{V} E(x,t) dx  = q^{*}(a,t)- q^{*}(b,t). }{claw}

We may easily see that if the model does not exhibit memory effects then identity (\ref{claw}) leads to classical one-phase Stefan problem. We state this result in the remark.
\begin{remark}
If the flux is defined by the Fourier law $q^{*}(x,t)= -  T_{x}(x,t)$, then (\ref{claw}) leads to the classical Stefan problem
\eqq{\ddt T(x,t)- T_{xx}(x,t)=0  \hd \m{ for  } \hd t>0 \hd \m{ and } \hd x\in (0,L)\setminus \{ s(t)\}, }{b2}
%\eqq{\dot s(t)=L(t),}{b3}
%where $L(t)$ is defined by
%\eqq{
%T_{x}^{\pm}(s(t),t) = \lim\limits_{y\rightarrow s(t)^{\pm}}T_{x}(y,t), \hd \hd L(t) =T_{x}^{+}(s(t),t)- T_{x}^{-}(s(t),t).
%}{a6d}
\eqq{\dot s(t) = - T_{x}^{-}(s(t),t)  \hd \m{ for  } \hd t>0,}{a6d}
where $s(t)$ is a interface and
\[
T_{x}^{-}(s(t),t) = \lim\limits_{\ep \rightarrow 0^{+}}T_{x}(s(t)-\ep,t).
\]
\end{remark}

%\section{Flux with the memory}
We assume that the flux is given by the Riemann-Liouville fractional derivative with respect to the time variable, i.e.
\eqq{q^{*}(x,t)= - \rlz \dtjmb T_{x}(x,t), }{memflux}
where
\[
\rlz \dtjmb T_{x}(x,t) = \frac{1}{\Gamma(\beta)} \ddt \int_{0}^{t} (t - \tau)^{\beta-1}T_{x}(x,\tau) d\tau, \hd \beta\in (0,1).
\]
In the next remark we give a formal idea why such form of the flux seems to be reasonable in the model exhibiting memory effects.
\begin{remark}
Let us denote by $s(t)$ the phase interface. We decompose the domain $\Om$ on the solid and liquid parts.
\[
\Omlt = (0, s(t)) \m{  - liquid}, \hd \hd \Omst = (s(t),L) \m{  - solid}.
\]
Let $V\ww \Om$ be arbitrary. Then, if we assume that $V=(a,b)$ and denote
\[
\vlt=\Omlt \cap V, \hd \hd \vst = \Omst \cap V
\]
then, (\ref{claw}) has a form
\eqq{\ddt \nk{ \int_{\vlt } (T(x,t)+ 1) dx   }  + \ddt \nk{ \int_{\vst} T(x,t) dx  } = \rlz \dtjmb  T_{x}(b,t) - \rlz \dtjmb T_{x}(a,t). }{clawmem}
Assuming that the temperature gradient is bounded with respect to time variable, after integrating with resect to time  we get
\[
\int_{\vlt } (T(x,t)+ 1) dx     +  \int_{\vst} T(x,t) dx = \int_{V_{l}(0) } (T(x,0)+ 1) dx     +  \int_{V_{s}(0)} T(x,0) dx
\]
\eqq{  +\frac{1}{\Gamma(\beta)}  \int_{0}^{t} (t- \tau )^{\beta-1} \nk{ T_{x}(b,\tau) - T_{x}(a,\tau)  } d \tau, }{d1}
i.e. the total enthalpy in $V$ at time $t$ is a sum of the initial enthalpy and the time-average of differences of local fluxes at the endpoints of $V$.

\end{remark}

\section{Derivation of the model}
In this section we derive (see (\ref{d8})-(\ref{d11})) the fractional Stefan model from the balance law (\ref{claw}) with the diffusive flux given by (\ref{memflux}). In order to do it rigorously we have to impose some regularity conditions on the phase interface $s$ and the temperature function $T$.
At first, we assume that $t^{*}$ is positive and
\[
s(t) \in AC[0,t^{*}], \hd  T_{x}(x,\cdot) \in L^{\infty}(U^{x}) \m{ for every } x \in \Omega,
\]
\begin{equation}
T_{x}(\cdot,t) \in AC[0,s(t)-\varepsilon] \m{ for every } \varepsilon > 0 \m{ and every } t \in (0,t^{*}),
\tag{A1}
\end{equation}
\[
T_{t}(\cdot, t)\in L^{1}(0,s(t)) \hd \m{ for each } t\in (0,t^{*}),
\]
where we denote
\[
Q_{s,t^{*}}=\{(x,t): \hd 0<x<s(t), \hd t\in (0,t^{*}) \},
\]
\[
U^{x}=\{t: \hd (x,t)\in Q_{s,t^{*}} \}
\]
and $AC$ denotes the space of absolutely continuous functions.

\no We equip our model with an initial condition:
\[
T(x,0)=T_{0}(x) \geq 0
\]
and the Dirichlet or Neumann boundary condition
\[
T(0,t)=T_{D}(t)\geq 0 \hd \m{ or } \hd T_{x}(0,t)=T_{N}(t) \leq 0.
\]
If $T_{0}$, $T_{D}\equiv 0$ or $T_{0}$, $T_{N}\equiv 0$, then we expect that $T\equiv 0$. Otherwise, we expect
\eqq{\dot s (t)>0,\tag{A2}}{c1}
i.e. we observe melting of solid. Since we consider one-phase problem, we have  $T(x,t)=0$ for $x\in \Omst$. Therefore, the flux is nonzero only in liquid part of the domain, i.e. in $Q_{s,t^{*}}$ and it is given by the formula
\eqq{q^{*}(x,t)= \left\{  \begin{array}{cll} - \rls \dtjmb  T_{x}(x,t)&   \m{ for }   & (x,t)\in Q_{s,t^{*}}, \\
0 &   \m{ for }   & (x,t)\not \in Q_{s,t^{*}},
\end{array}
 \right. }{memfluxe}
where
\eqq{\rls \dtjmb  T_{x}(x,t) = \left\{ \begin{array}{lll}  \frac{1}{\Gamma(\beta)} \ddt  \int_{0}^{t} (t- \tau)^{\beta-1}  T_{x}(x,\tau) d \tau   \hd & \m{ for  } \hd & x\leq s(0) \\ \frac{1}{\Gamma(\beta)} \ddt  \int_{s^{-1}(x)}^{t} (t- \tau)^{\beta-1}  T_{x}(x,\tau) d \tau   \hd & \m{ for  } \hd & x> s(0). \\ \end{array}  \right.}{c2}
We recall the definitions of fractional integral ${}_{0}I_{t}^{\beta}$ and fractional Caputo derivative $\cabt$ for a later use
\[
{}_{0}I_{t}^{\beta} f(x,t)= \frac{1}{\Gamma(\beta)}\int_{0}^{t} (t- \tau)^{\beta-1}f(x, \tau) d\tau,
\hd \hd
\cabt f(x,t)= \frac{1}{\Gamma(1-\beta)}\int_{0}^{t} (t- \tau)^{-\beta}f_{t}(x, \tau) d\tau.
\]

\no We note that since we consider one-phase Stefan problem the temperature in the solid vanishes. This together with (\ref{a1}) leads to the following form of equality  (\ref{claw})
\eqq{\ddt \nk{ \int_{\vlt } T(x,t)+ 1 dx   }   = -q^{*}(b,t) + q^{*}(a,t). }{clawmems}

In order to derive the system of equations from (\ref{clawmems}) we fix a positive time $t \in (0,t^{*})$ and we consider the conservation of energy on a subset $V$ of the domain at time $t$. We will consider two cases.
\begin{itemize}
\item
If $V=(a,b)\ww (0,s(0))$,  then from (A2) we have $V \ww (0,s(t))$ for each $t\in (0,t^{*})$ and (\ref{clawmems}) gives
\[
\ddt \nk{ \int_{V} T(x,t)+1 dx}=\rlz \dtjmb  T_{x}(b,t) - \rlz \dtjmb T_{x}(a,t)
\]
hence,
\[
  \int_{V} \ddt T(x,t) dx=\rlz \dtjmb  T_{x}(b,t) - \rlz \dtjmb T_{x}(a,t).
\]
We apply the fractional integral $\intfjmb$ with respect to the time variable to both sides of the identity and with a use of assumption (A1) we arrive at
\[
\int_{V} \cabt T(x,t) dx = T_{x}(b,t) - T_{x}(a,t)
\]
and after using the fundamental theorem of calculus we obtain
\[
\int_{V} [ \cabt T(x,t) -  T_{xx}(x,t)]dx =0.
\]
Since $V \ww (0,s(0)) $ is arbitrary, we get
\eqq{\cabt T(x,t) -  T_{xx}(x,t) =0 \hd \hd \m{ for } \hd (x,t)\in (0,s(0))\times (0,t^{*}).}{a2es}
\item
If $V=(a,b)$, where $s(0)<a<s(t)<b$, then (\ref{clawmems}) has a form
\[
\ddt \nk{ \int_{a}^{s(t)} T(x,t)+1 dx} = q^{*}(a,t) = - \frac{1}{\Gamma(\beta)} \ddt \int_{\sma}^{t} (t-\tau)^{\beta-1}T_{x}(a,\tau) d \tau.
\]
Differentiating  the integral on the left hand side leads to
\[
\int_{a}^{s(t)} \ddt T(x,t) dx+\dot s(t)[T(s(t),t)+1] =  - \frac{1}{\Gamma(\beta)} \ddt \int_{\sma}^{t} (t-\tau)^{\beta-1}T_{x}(a,\tau) d \tau.
\]
Using $T(s(t),t)=0$ and applying operator $\frac{1}{\Gamma(1-\beta)}\int_{\sma}^{t} (t-\tau)^{-\beta}\cdot d\tau$ we obtain
\[
\frac{1}{\Gamma(1-\beta)} \int_{\sma}^{t}(t-\tau)^{-\beta} \int_{a}^{s(\tau)} \ddta T(x,\tau) dxd\tau+\frac{1}{\Gamma(1-\beta)} \int_{\sma}^{t}(t-\tau)^{-\beta} \dot s(\tau)d\tau
\]
\eqq{
=  - \frac{1}{\Gamma(\beta)}\frac{1}{\Gamma(1-\beta)} \int_{\sma}^{t}(t-\tau)^{-\beta} \ddta \int_{\sma}^{\tau} (\tau-p)^{\beta-1}T_{x}(a,p) dp d \tau.
}{dj}
Let us denote
\eqq{\cs \dtb  T(x,t) = \left\{ \begin{array}{lll}  \frac{1}{\Gamma(1-\beta)}   \int_{0}^{t} (t- \tau)^{-\beta}  \ddta T(x,\tau) d \tau   \hd & \m{ for  } \hd & x\leq s(0) \\ \frac{1}{\Gamma(1-\beta)}   \int_{s^{-1}(x)}^{t} (t- \tau)^{-\beta} \ddta T(x,\tau) d \tau   \hd & \m{ for  } \hd & x> s(0). \\ \end{array}  \right.}{d2}
If we apply the Fubini theorem to the first and third integral in (\ref{dj}) and make use of the assumption (A1) we arrive at the identity
\eqq{
\int_{a}^{s(t)} \cs \dtb  T(x,t)dx+ \frac{1}{\Gamma(1-\beta)} \int_{\sma}^{t}(t-\tau)^{-\beta} \dot s(\tau)d\tau = - T_{x}(a,t).
}{dc}
Applying the substitution $\tau = s^{-1}(x)$ we get
\[
\frac{1}{\Gamma(1-\beta)} \int_{\sma}^{t}(t-\tau)^{-\beta} \dot s(\tau)d\tau = \frac{1}{\Gamma(1-\beta)} \int_{a}^{s(t)} (t-s^{-1}(x))^{-\beta} dx.
\]
We expect that $T_{x}$ may admit singular behaviour near the phase change point. Thus, we proceed very carefully and fix $\ep>0$ such that $a<s(t)-\ep$. Then, by (A1) we have
\[
- T_{x}(a,t) = \int_{a}^{s(t)-\ep} T_{xx}(x,t)dx - T_{x}(s(t)-\ep,t).
\]
Making use of this results in (\ref{dc}) we obtain
\[
\int_{a}^{s(t)-\ep} \nk{ \cs \dtb  T(x,t)dx-T_{xx}(x,t)+\frac{1}{\Gamma(1-\beta)} (t-s^{-1}(x))^{-\beta} }dx
\]
\eqq{=-\int_{s(t)-\ep}^{s(t)} \nk{ \cs \dtb  T(x,t)dx+\frac{1}{\Gamma(1-\beta)} (t-s^{-1}(x))^{-\beta} }dx -T_{x}(s(t)-\ep,t).  }{d3}
Let us choose arbitrary $\ti{a}$ such that $s(0)<\ti{a}<a$, then we get
\[
\int_{\ti{a}}^{s(t)-\ep} \nk{ \cs \dtb  T(x,t)dx-T_{xx}(x,t)+\frac{1}{\Gamma(1-\beta)} (t-s^{-1}(x))^{-\beta} }dx
\]
\eqq{=-\int_{s(t)-\ep}^{s(t)} \nk{ \cs \dtb  T(x,t)dx+\frac{1}{\Gamma(1-\beta)} (t-s^{-1}(x))^{-\beta} }dx -T_{x}(s(t)-\ep,t).  }{d33}
Subtracting the sides of (\ref{d3}) and (\ref{d33}) we arrive at
\eqq{\int_{\ti{a}}^{a} \nk{ \cs \dtb  T(x,t)dx-T_{xx}(x,t)+\frac{1}{\Gamma(1-\beta)} (t-s^{-1}(x))^{-\beta} }dx =0 }{d12}
for arbitrary $a, \ti{a} \in (s(0),s(t)-\ep)$ hence, we may deduce that
\eqq{ \cs \dtb  T(x,t)dx-T_{xx}(x,t)+\frac{1}{\Gamma(1-\beta)} (t-s^{-1}(x))^{-\beta} = 0 \hd \m{ for } \hd x\in (s(0), s(t)). }{d6}
\no We impose another regularity assumption concerning $s$. Namely,
\eqq{t^{1-\beta}\dot s(t) \in L^{\infty}(0,t^{*}). \tag{A3}}{e1}
Then
\eqq{ \lim_{\ep \rightarrow 0^{+}} \int_{s(t)-\ep}^{s(t)}(t-s^{-1}(x))^{-\beta}dx=0. }{e2}
Next, we assume that 
\eqq{ \forall t \in (0,t^{*}) \hd \exists \ep_{0}>0, \hd \exists a>\frac{1}{1-\beta} \hd \m{ such that } \hd    \int_{s(t)-\ep_{0}}^{s(t)} \int_{s^{-1}(x) }^{t}  |T_{t}(x,\tau)|^{a} dx d\tau <\infty.\tag{A4}}{e3}
Then, from the above assumption, for $\ep\in (0,\ep_{0})$ we have
\[
\left| \int_{s(t)-\ep}^{s(t)}  \cs \dtb  T(x,t)dx \right| 
\]
\[
\leq \frac{1}{\Gamma(1- \beta)}  \left(  \int_{s(t)-\ep_{0}}^{s(t)} \int_{s^{-1}(x) }^{t}  |T_{t}(x,\tau)|^{a} dx d\tau  \right)^{1/a}  \left| \int_{s(t)-\ep}^{s(t)}  \int_{s^{-1}(x) }^{t}  (t- \tau)^{-\frac{a\beta}{a-1}} d\tau dx \right|^{\frac{a-1}{a}}
\]
\[
\]
%if $\ep \rightarrow 0^{+}$, because $s(t)$ is continuous and
%\[
%\int_{s(t)-\ep}^{s(t)}  \int_{s^{-1}(x) }^{t}  (t- \tau)^{-\beta} d\tau dx = \int_{s^{-1}(s(t)-\ep)}^{t} \int_{s(t)-\ep}^{s(\tau) } (t- \tau)^{-\beta} dx d\tau
%\]
%\[
%= \int_{s^{-1}(s(t)-\ep)}^{t} (t- \tau)^{-\beta} [s(\tau) - s(t)+\ep] d\tau \leq \frac{\ep}{1- \beta} [t - s^{-1}(s(t)-\ep)]\rightarrow 0.
%\]
hence, we get
\eqq{\left| \int_{s(t)-\ep}^{s(t)}  \cs \dtb  T(x,t)dx \right|\rightarrow 0, \hd \m{ if } \hd \ep \rightarrow 0^{+}.}{e4}
We note that (\ref{d33}) and (\ref{d6}) lead to
\[
0=-\int_{s(t)-\ep}^{s(t)} \nk{ \cs \dtb  T(x,t)dx+\frac{1}{\Gamma(1-\beta)} (t-s^{-1}(x))^{-\beta} }dx -T_{x}(s(t)-\ep,t).
\]
Making use of (\ref{e2}) and (\ref{e4}) we obtain
\eqq{\lim_{\ep \rightarrow 0^{+}} T_{x}(s(t)-\ep,t) = 0. }{e5}
Finally, we have obtained the following system
\eqq{ \cs \dtb  T(x,t)dx-T_{xx}(x,t)= \left\{  \begin{array}{cll} 0 & \m{ for } & x<s(0) \\  - \frac{1}{\Gamma(1-\beta)} (t-s^{-1}(x))^{-\beta}  &  \m{ for } &  x\in (s(0), s(t)) \end{array}\right.}{d8}
with the boundary conditions
\eqq{T(s(t),t)=0, \hd \hd T_{x}(s(t),t)=0,}{d9}
\eqq{T(0,t)=T_{D}(t)\geq 0 \hd \m{ or } \hd T_{x}(0,t)= T_{N}(t)\leq 0,}{d10}
and the initial condition
\eqq{T(x,0)=T_{0}(x) \geq 0.}{d11}

\end{itemize}

\end{document}